\documentstyle[prl,aps]{revtex} 
\draft
\begin{document}
\twocolumn[\hsize\textwidth\columnwidth\hsize\csname
@twocolumnfalse\endcsname

\title{Classical and Quantum Gravity on Conformal Superspace}

\author{Julian B. Barbour\thanks{email: jbarbour@online.rednet.co.uk}}
\address{College Farm, South Newington, Banbury, Oxfordshire, OX15 4JG, UK} 
\author{Niall \'O Murchadha\thanks{email: niall@ucc.ie}}
\address{Physics Department, University College, Cork, Ireland} 
\date{\today}
\maketitle 
\begin{abstract}
   The four-dimensional gauge group of general relativity corresponds to
arbitrary coordinate transformations on a four-manifold (spacetime). Theories
of gravity with a dynamical structure remarkably like Einstein's
theory can be obtained on the basis of a four-dimensional gauge group of
arbitrary coordinate and {\it conformal} transformations of riemannian metrics
defined on a three-manifold. This new symmetry is more restrictive and hence
more predictive. Many of the difficulties that have plagued the canonical
quantization of general relativity seem to vanish.

\end{abstract}

\vskip2pc]
Let us consider a number of spaces: {\it Riem}, the set of
all riemannian three-metrics on a given compact manifold; {\it superspace},
obtained from {\it Riem} by identifying as three-geometries
all three-metrics that are related by coordinate
transformations; and {\it conformal superspace}, ($CS$),  obtained from
superspace by identifying all three-geometries whose metrics are related by
conformal rescalings $g_{ab} \rightarrow \omega^4 g_{ab}$ where $\omega$ is
an arbitrary strictly positive function. One can regard the conformal factor
as a fourth `coordinate' on three-space. 
We also consider {\it restricted conformal superspace}, ($CS^*$), in
which only conformal three-geometries with the same volume are identified.

In canonical general relativity \cite{adm} one is given a pair $\{g_{ab},
\pi^{ab}\}$ where $g_{ab}$ is a riemannian three-metric and $\pi^{ab}$, the
conjugate momentum, is a symmetric three-tensor-density. These 
must satisfy the four constraints
\begin{equation}
\pi^{ab}_{;b} = 0;\hskip 1cm gR = \pi^{ab}\pi_{ab} - {1 \over 2}(tr
\pi)^2, \label{adm}
\end{equation}
where $R$ is the scalar curvature of $g_{ab}$. 

There are twelve degrees of freedom per space point in the pair $\{g_{ab},
\pi^{ab}\}$ but three of them represent the three coordinates 
 and we also need to include the four constraints. Hence the initial data
really have five degrees of freedom. Four represent
the true gravitational degrees of freedom while the fifth is kinematical and
represents the freedom to imbed the spacelike slice in the spacetime.

York has cogently argued \cite{jwy} that the four  dynamical degrees of
freedom are coded into the conformal geometry of the spacelike slice and that
the configuration space of gravity should be conformal superspace.
This {\it
almost} works in general relativity. The six components of the metric reduce to
two when one subtracts off  one conformal and three coordinate degrees of
freedom. Also any symmetric tensor,
$A_{ab}$, can be uniquely
decomposed\cite{jwy2}
\begin{eqnarray}
A_{ab} &=& A^{TT}_{ab} + (LW)_{ab} + \theta g_{ab}\nonumber\\ 
&=& A^{TT}_{ab} +
(KW)_{ab} +
\bar \theta g_{ab}
\end{eqnarray}
where $(LW)_{ab} = W_{a;b} + W_{b;a} - {2\over 3} W^k{}_{;k}g_{ab}$ is the
conformal killing form of a vector $W^k$; $(KW)_{ab} =  W_{a;b} + W_{b;a}$ is
the killing form of the same vector; 3$\theta = tr A = g_{ab}A^{ab}$; and
$\bar \theta =
\theta  - {2 \over 3}W^k{}_{;k}$. $A^{TT}$ is both tracefree and divergencefree.
TT tensors are conformally covariant: if $ A^{TT}_{ab}$ is TT with respect to a
given metric $g_{ab}$, then $\omega^{-2} A^{TT}_{ab}$ is TT with respect the
conformally related metric $g'_{ab} = \omega^4 g_{ab}$. Thus any TT tensor
represents a tangent vector in conformal superspace and clearly has two degrees
of freedom per space point. In general relativity the standard approach
is to choose the trace of the extrinsic curvature as the fifth degree of
freedom. This breaks the conformal invariance.

We intend to take the  conformal
structure seriously and construct a new theory of gravity. We find a
conformally invariant action which gives a measure (a `metric') on conformal
superspace. The solutions of the theory will be a family of unique geodesics in
the configuration space. Since we start with a Jacobi action, the
geodesic will be a parametrised curve. The parameter itself has no intrinsic
meaning. It is remarkable how closely these curves in conformal superspace
match curves in superspace which represent solutions of the Einstein
equations.

In 1962 Baierlein, Sharp and Wheeler (BSW) \cite{bsw} constructed a Jacobi
action for G.R. It was of the form
\begin{equation}
I = \int d\lambda \int \sqrt{g} \sqrt{R} \sqrt{T} d^3x,
\end{equation}
where the `kinetic energy' $T$ is 
\begin{eqnarray}
T &=& (g^{ac}g^{bd} - g^{ab}g^{cd})\nonumber\\& &\left({\partial g_{ab} \over
\partial
\lambda} - (KW)_{ab}\right)\left({\partial g_{cd} \over \partial\lambda} -
(KW)_{cd}\right).
\end{eqnarray}
This action reproduces the standard Einstein equations in the thin-sandwich
form with lapse $N = \sqrt{T/4R}$.

 Barbour and Bertotti \cite{bb} realised
that this action could be constructed naturally by a `best matching'
procedure. One picks two nearby metrics $g_{ab}$ and $g_{ab} + \delta g_{ab}$
and tries to measure a separation between them while allowing for an arbitrary
coordinate transformation on the second metric and simultaneously using the
`potential energy' term
$\sqrt{R}$ as a weighting. Hence one minimizes the action over all vectors
$W^a$. In fact, BSW actually had two coordinate transformations: one 
generated by $W^a$, the other which is implemented 
by the action being a geometric scalar.

The simplest conformalization of the BSW action is
\begin{equation}
I = \int d\lambda {\int \sqrt{g}\phi^4 \sqrt{R - {8\nabla^2\phi \over \phi}}
\sqrt{T} d^3x \over  V(\phi)^{{2 \over 3}}},
\label{5}
\end{equation}
where $V(\phi) = \int \phi^6 \sqrt{g}d^3x $ and the new  $T$ is
\begin{eqnarray}
&T& = (g^{ac}g^{bd} - Ag^{ab}g^{cd})\nonumber\\& &\left({\partial g_{ab} \over
\partial\lambda} - (LW)_{ab} - \theta g_{ab}\right) \left({\partial g_{cd} \over
\partial\lambda} - (LW)_{cd} - \theta g_{cd}\right),
\label{T}
\end{eqnarray}
with $A$ an (as  yet) arbitrary constant. We do not restrict ourselves to the 
DeWitt form but rather allow a more general
supermetric in the theory . 
  The
denominator must be chosen to be of the same degree in $\phi$ as the numerator
so that one cannot make the action vanish by a simple scaling on $\phi$.
Other similar conformally invariant actions
can be found by changing  the power of $R$ in the
numerator and appropriately changing the denominator.

 As in Ref.\cite{bb} we compare two nearby metrics; however,
in addition to the coordinate transformations, the new scalar,
$\theta$, in the kinetic energy allows us to make an arbitrary conformal
rescaling between the slices while the second
function, $\phi$, allows an overall conformal rescaling. 

 Both the numerator and denominator are conformally invariant. Choose an
arbitrary positive function
$\omega(x^i, \lambda)$ and consider the following mapping
\begin{eqnarray}
g'_{ab} &=& \omega^4g_{ab}, {\partial g_{ab}'\over \partial \lambda} = 
\omega^4{\partial g_{ab}\over \partial \lambda} + 4\omega^3{\partial
\omega \over \partial \lambda}g_{ab},\nonumber\\ 
\phi' &=& {\phi\over\omega}, W^a{}' = W^a, \theta' = \theta +
4\omega^{-1}{\partial
\omega \over \partial \lambda}.\label{7}
\end{eqnarray}
This gives
\begin{eqnarray}
R'- {8\nabla'{}^2 \phi' \over \phi'} &=& \omega^{-4}(R - {8\nabla^2 \phi
\over \phi}),\nonumber\\ (L'W')_{ab} &=& \omega^4(LW)_{ab},\hskip 0.5cm T' = T,
\nonumber\\
\sqrt{g'}\phi'{}^4\sqrt{R'- {8\nabla'{}^2 \phi' \over \phi'}}\sqrt{T'} &=&
\sqrt{g}\phi^{4}\sqrt{R - {8\nabla^2
\phi
\over \phi}}\sqrt{T}.
\end{eqnarray}
We find the constraints of the theory by varying with respect to $\phi$, $W$,
and $\theta$. All the
equations will be conformally invariant. Only the TT part of $\partial g/
\partial \lambda$ contributes to the action; however, after we solve the
constraints, $\partial g/\partial
\lambda - (LW) - \theta g$  will not be just the
TT part of  $\partial g/\partial \lambda$, it will have a vector part arising
from the fact that the potential energy is not constant.

The momentum conjugate to $g_{ab}$, found by varying the action
with respect to  $\partial g/\partial \lambda$, is
\begin{eqnarray}
\pi^{ab} &=& {\sqrt{g}\phi^4 \sqrt{R - {8\nabla^2\phi \over \phi}}
\over \sqrt{T}V(\phi)^{2\over 3}}\nonumber\\
& &(g^{ac}g^{bd} -
Ag^{ab}g^{cd})\left({\partial g_{cd} \over
\partial\lambda} - (LW)_{cd} - \theta g_{cd}\right).\label{9}
\end{eqnarray}
Note that we go from a displacement
to a direction  in {\it Riem} because $\sqrt{T}$ is the norm of the $\partial
g/\partial
\lambda$ term and so 
\begin{equation}
\pi^{ab}\pi_{ab} - {A \over 3A - 1}(tr \pi)^2 = {g \phi^8 \over V(\phi)^{{4
\over 3}}}\left(R - {8\nabla^2\phi \over \phi}\right).\label{10}
\end{equation}
which is a reparametrisation identity arising directly from Eq.(\ref{9}).
When we vary the action w.r.t. $W^a$ and $\theta$ we get the striking
 result that $\pi^{ab}$ is TT and thus a direction in conformal
superspace. In other words
\begin{equation}
\pi^{ab}{}_{;b} = 0, \hskip 1cm tr\pi = 0.
\end{equation}
This shows that the coefficient $A$ in the supermetric can be set to zero; the
kinetic energy is positive; and the reparametrisation identity reduces to
\begin{equation}
\pi^{ab}\pi_{ab} \equiv {g \phi^8 \over V(\phi)^{{4 \over 3}}}\left(R -
{8\nabla^2\phi \over \phi}\right).\label{12}
\end{equation}

We next consider the variation w.r.t. $\phi$. Effectively we are minimizing the
BSW action on a fixed metric and further minimizing it by making an
overall conformal transformation as defined by Eq.(\ref{7}). The
minimizing $\phi$ is the conformal factor that brings us to the minimizing
metric. The equation is
\begin{equation}
{\sqrt{T}(\phi^3R - 7\phi^2\nabla^2\phi) \over \sqrt{R - {8\nabla^2\phi\over
\phi}}} - \nabla^2[{\sqrt{T} \phi^3 \over \sqrt{R - {8\nabla^2\phi\over
\phi}}}] = {\bar{C}\phi^5 \over V(\phi)^{1 \over 3}}.
\label{13}
\end{equation}
This is a
conformally invariant eigenvalue equation. The eigenvalue $\bar C$ arises from
the variation of the denominator and it is what prevents $\phi \equiv 0$ from
being a solution. We call this the energy norm equation.

Implementing $tr\pi = 0$ is trivial: one subtracts off
the trace of $\partial g/\partial\lambda$. The energy norm equation and 
setting $\pi^{ab}{}_{;b} = 0$ are more difficult: one finds a set of four
nonlinear coupled equations. These are analagous to the thin sandwich equations
of general relativity
\cite{rb}. We presume that they can be solved for a range of  $\{g_{ab},
\partial g_{ab}/\partial \lambda$\}.

The problem of solving these equations can be avoided. 
 We construct the hamiltonian version of conformal gravity
and it is, as in general relativity, better posed than the
thin sandwich version.
The initial data now consist of a metric and a TT tensor density, $\{ g_{ab},
V(\phi)^{2 \over 3}\pi^{ab}_{TT}\}$, these should be thought of as a point and direction in
conformal superspace. Since $\pi^{ab}$ (and not $\partial g_{ab}/\partial
\lambda$) is now the basic variable the reparametrisation identity,
Eq.(\ref{10}), becomes an equation and is solved for a positive
$\phi = \phi_s$. This $\phi_s$ is substituted into the energy norm equation,
Eq.(\ref{13}), which, in turn, is solved for
$T$. We can (if we wish) use the solution, $\phi_s$, of the reparametrisation
equation as a conformal factor to simplify matters.  We call the system in this
`simplest' state the best-matched representation. The solution of the
reparametrisation identity, in the best-matched representation, following
Eq.(\ref{7}), is
$\phi'_s =\phi_s/\omega =
\phi_s/\phi_s \equiv 1$. Thus the energy norm equation, in the best matched
representation, reduces to
\begin{equation}
\sqrt{T \over R}R - \nabla^2 \sqrt{T\over R} = {C \over V},
\label{14}
\end{equation}
with the eigenvalue $C$ satisfying $C = \int\sqrt{gRT}d^3x$, the numerator of
the Lagrangian. This equation is homogeneous and thus the solution
has an undetermined overall scale factor. It is this scale factor which allows
the global reparametrisation of the solution curve in configuration space.

The equation for the time derivative of $tr\pi$ in general relativity is
\begin{equation}
{\partial \over \partial t}\left({tr \pi \over \sqrt{g}}\right) = 2RN + {N(tr
\pi)^2 \over g}  - 2\nabla^2N  + \left({tr \pi \over \sqrt{g}}\right)_{;a}N^a.
\end{equation}
Thus the equation for the lapse function which generates
constant mean curvature slices at $tr\pi = 0$ is
\begin{equation}
RN - \nabla^2N = C_1,
\end{equation}
essentially the same as Eq.(\ref{14}).
In the best-matched representation the reparametrisation identity,
Eq.(\ref{12}), looks just like the hamiltonian constraint of general relativity
at maximal expansion if we multiply the conformal gravity momentum by
$V^{2\over 3}$ . Further, if we compare Eq.(\ref{9}) to the definition of the
momentum in canonical general
relativity, it is clear that the natural relationship is $2N = \sqrt{T/R}$,
just as in BSW.

The lagrangian equation (or the second hamiltonian equation) is $\partial
\pi^{ab}/\partial \lambda = \partial L/\partial g_{ab}$. Thus the dynamical
equations in the best-matched representation are 
\begin{eqnarray}
{\partial g_{ab} \over \partial \lambda} &=& \sqrt{T \over gR} V^{2 \over
3}\pi_{ab}\\
V^{2 \over 3} {\partial \pi^{ab} \over \partial \lambda} &=& {1 \over 2}
\sqrt{gRT}g^{ab} - {1 \over 2}\sqrt{gT \over R} R^{ab} + {1 \over
2}\left(\sqrt{gT
\over R}\right)^{;ab}\nonumber\\ &-& {1 \over 2}\nabla^2\sqrt{gT \over R}g^{ab}
-\sqrt{T\over gR}\pi^{ac}\pi^b{}_c - {\sqrt{g}C \over 3V}g^{ab}.
\end{eqnarray}
The initial data consist of a pair $\{g_{ab}, \pi^{ab}\}$ which satisfy the
three constraints
\begin{equation}
\pi^{ab}{}_{;b} = 0, \hskip 0.5cm tr\pi = 0, \hskip 0.5cm V^{4 \over 3}
\pi^{ab}\pi_{ab} = gR,
\end{equation}
and the function $T$ must satisfy Eq.(\ref{14}). It is easy to
show that the evolution equations preserve the constraints. Therefore they 
generate a unique curve in {\it Riem} which stays in the best-matched
representation. We can add to each of the evolution equations a Lie derivative
with respect to an arbitrary shift. This
will give us a curve in superspace that remains in the best-matched
representation. This is a representative of our desired curve in CS. If we
substitute
$\{g, {\partial g/\partial \lambda}\}$ from this curve into the original
action, Eq.(\ref{5}), the best-matching procedure will give us $\phi = 1,
W^i = 0, \theta = 0$.

It is remarkable that a scale-free theory nevertheless leads, through
best-matching minimization, to a metric with scale. It was shown in \cite{bb}
how local proper time emerges through best-matching on superspace. Now local
lengths emerge from scale-free best-matching on conformal superspace. The
determination of the full metric via the Lichnerowicz equation, which is
essentially our Eq.(\ref{12}), as the final step in the York procedure has
usually been regarded as a useful construct rather than something fundemental.
Our work shows that it is natural and inevitable in conformal gravity.

Let us consider a cosmology which satisfies the vacuum Einstein equations and
goes from a big bang to a big crunch. This will have a moment of maximum
expansion. At this point we will have initial data for both general relativity
and conformal gravity using the relationship $\pi^{ab}_{CG} =V^{2 \over 3}
\pi^{ab}_{GR}$. Let us propagate the Einstein initial data in the constant mean
curvature gauge and the conformal gravity data in the best-matched
representation. We find that initially $N$ is proportional to $\sqrt{T/R}$
and can arrange
\begin{equation}
\left[{\partial g_{ab} \over \partial t}\right]_{GR} = \left[{\partial g_{ab}
\over \partial \lambda}\right]_{CG};  V^{2 \over 3}\left[{\partial
\pi^{ab}
\over \partial t}\right]_{GR} = \left[{\partial \pi^{ab} \over \partial
\lambda}\right]_{CG}.
\end{equation}
We cannot maintain this matching at higher orders, because the GR momentum
develops a trace, but we can certainly match to the next order by using a
conformal rescaling to move the conformal gravity curve out of the
best-matched representation.

It is easy to work out the hamiltonian. It is

\begin{eqnarray}
H &=& {\sqrt{T}V(\phi)^{2 \over 3} \over 2\sqrt{g}\phi^4\sqrt{R - {8\nabla^2\phi
\over \phi}}}\left[\pi^{ab}\pi_{ab} - {g\phi^8\left(R - {8\nabla^2\phi
\over \phi}\right) \over V(\phi)^{4 \over 3}}\right]\nonumber\\& & 
-2W_a\pi^{ab}{}_{;b} +
\bar \theta tr \pi.\label{20}
\end{eqnarray}
It is the sum of the three constraints with lagrange multipliers. We have
four quantities without associated momenta, $W^a$, $\theta$, $T$, and $\phi$.
When we vary w.r.t. the first three we get the three constraints, when we vary
w.r.t. $\phi$ we get the energy norm equation, Eq.(\ref{13}).

This analysis works in the case where the manifold is compact without
boundary. In the asymptotically flat case we cannot use the volume as
denominator. The obvious solution is to use the numerator as the action and
to control the conformal factor by the requirement that $\phi \rightarrow 1$ at
infinity. This means that the energy norm equation is no longer an
eigenvalue equation, and Eq.(\ref{14}) becomes
\begin{equation}
\nabla^2\sqrt{T \over R} - \sqrt{T \over R} R = 0.
\end{equation} 
This is the maximal slicing equation. Thus solutions of the
vacuum Einstein equations in the maximal gauge (as curves in superspace) agree
exactly with solutions of the conformally invariant equations in the
best matched representation. Therefore conformal gravity should pass all the
standard tests. The hamiltonian will look just like Eq.(\ref{20}) except that
the $V(\phi)$ is omitted. This is nondifferentiable and
the usual surface terms will have to be added to control the
integration by parts\cite{rt}. The positive energy theorem\cite{sy}
continues to hold.

There are solutions of the vacuum Einstein equations which are not linked to
solutions of the conformal equations. These are the solutions in GR which do
not have a maximal slice such as cosmological solutions which
expand forever. The reparametrisation identity demands that the scalar curvature
be positive. This
 severely restricts the possible topologies\cite{w}. Thus we
have some form of topological censorship.

Applying the standard canonical quantization procedure to this conformal theory
is quite straightforward: the reparametrisation identity converts into a
Wheeler-DeWitt equation and we get a time independent Schr\"odinger
equation which gives us a probability distribution on conformal superspace. The
other constraints act on the wavefunction to guarantee both
coordinate independence and conformally invariance. The time independence also
is natural: there is no time in the classical theory so why should there be one
in the quantum theory? Finally, the supermetric is positive definite: there are
none of the negative energy modes that bedevil the standard Wheeler-DeWitt
equation.

In light
of the success of this programme, it is certainly worth trying to
cast general relativity as a theory with a restricted conformal invariance.
Removing the denominator in the action Eq.(\ref{5}) and replacing it with the
term $ +\xi[V(\phi) - V_0]$, where $\xi$ is a lagrange
multiplier, has essentially no effect on the system. Both the
energy norm equation and the hamiltonian equations are
unchanged except that the $V$'s drop out. In this new version,  we
are making a restricted overall conformal transformation, one which
preserves the total volume. If we restrict the conformal
rescaling between the nearby metrics in the same fashion, we
replace $\theta$ by $b^a{}_{;a}$, where $b^a$ is an arbitrary vector
field. The constraint that arises from varying $b^a$ is $\nabla_a(tr \pi) =
0$, the constant mean curvature condition. In addition, we need to return to the
original form of the BSW supermetric, i.e., $g^{ac}g^{bd} - g^{ab}g^{cd}$.
Then, in the best matched representation, when $\phi \equiv 1$, the system is
identical to general relativity in the CMC gauge. Let us stress that
this theory is not quite as simple as it seems: it has a very
complicated gauge group because the kinetic energy
is not conformally invariant.

There are a number of obvious ways of generalizing this work, It is clear that
adding a constant to the scalar curvature term in the action is equivalent to
adding a cosmological constant to the theory. It would be interesting to couple
in matter; an obvious place to start is with fields which have their own
conformal invariance as in \cite{imy}. The other conformal theories with
different powers of the scalar curvature also need to be investigated.

\acknowledgements{This work has been partially supported by the Forbairt grant 
SC/96/750. We wish to thank Domenico Giulini, Ted Jacobson, Claus Kiefer, Karel
Kucha\u{r}, and Lee Smolin for helpful comments. }

\end{document}